\documentclass[aps, prapplied, superscriptaddress, amsmath,amssymb, reprint]{revtex4-2}
\usepackage{graphicx}
\usepackage{dcolumn}
\usepackage{bm}
\usepackage{tabularx}

\usepackage[utf8]{inputenc}
\usepackage[T1]{fontenc}
\usepackage{newtxtext,newtxmath} 
\usepackage{color, soul}
\usepackage{siunitx}
\usepackage{soul,xcolor}
\usepackage{comment}
\usepackage{textcomp}
\graphicspath{ {Figures/} }

\newcommand{\ba}{\begin{eqnarray}}
\newcommand{\ea}{\end{eqnarray}}

\begin{document}
\title{Crystalline superconductor-semiconductor Josephson junctions for compact superconducting qubits}
\author{Jesse Balgley}
\altaffiliation{These authors contributed equally to this work.}
\affiliation{Department of Mechanical Engineering, Columbia University, New York, NY 10027, USA}
\author{Jinho Park}
\altaffiliation{These authors contributed equally to this work.}
\affiliation{Department of Mechanical Engineering, Columbia University, New York, NY 10027, USA}
\author{Xuanjing Chu}
\altaffiliation{These authors contributed equally to this work.}
\affiliation{Department of Applied Physics and Mathematics, Columbia University, New York, NY 10027, USA}
\author{Ethan G.~Arnault}
\affiliation{Department of Electrical Engineering and Computer Science, Massachusetts Institute of Technology, Cambridge, MA 02139}
\author{Martin V.~Gustafsson}
\affiliation{RTX BBN Technologies, Quantum Engineering and Computing Group, Cambridge, MA 02138, USA}
\author{Leonardo Ranzani}
\affiliation{RTX BBN Technologies, Quantum Engineering and Computing Group, Cambridge, MA 02138, USA}
\author{Madisen Holbrook}
\affiliation{Department of Physics, Columbia University, New York, NY 10027, USA}
\author{Yangchen He}
\affiliation{Department of Materials Science and Engineering, University of Wisconsin--Madison, Madison, WI 53706, USA}
\author{Kenji Watanabe}
\affiliation{National Institute for Materials Science, 1-2-1 Sengen, Tsukuba, Ibaraki 305-0044, Japan}
\author{Takashi Taniguchi}
\affiliation{National Institute for Materials Science, 1-2-1 Sengen, Tsukuba, Ibaraki 305-0044, Japan}
\author{Daniel Rhodes}
\affiliation{Department of Materials Science and Engineering, University of Wisconsin--Madison, Madison, WI 53706, USA}
\affiliation{Department of Physics, University of Wisconsin--Madison, Madison, WI 53706, USA}
\author{Vasili Perebeinos}
\affiliation{Department of Electrical Engineering, University at Buffalo, The State University of New York, Buffalo, New York 14260, USA}
\author{James Hone}
\affiliation{Department of Mechanical Engineering, Columbia University, New York, NY 10027, USA}
\author{Kin Chung Fong}%
\altaffiliation{Present address:~k.fong@northeastern.edu, Northeastern University}
\affiliation{RTX BBN Technologies, Quantum Engineering and Computing Group, Cambridge, MA 02138, USA}
\date{\today}

\begin{abstract}
The narrow bandgap of semiconductors allows for thick, uniform Josephson junction barriers, potentially enabling reproducible, stable, and compact superconducting qubits. We study vertically stacked van der Waals Josephson junctions with semiconducting weak links, whose crystalline structures and clean interfaces offer a promising platform for quantum devices. We observe robust Josephson coupling across 2--12 nm (3--18 atomic layers) of semiconducting WSe$_2$ and, notably, a crossover from proximity- to tunneling-type behavior with increasing weak link thickness. Building on these results, we fabricate a prototype all-crystalline merged-element transmon qubit with transmon frequency and anharmonicity closely matching design parameters. We demonstrate dispersive coupling between this transmon and a microwave resonator, highlighting the potential of crystalline superconductor-semiconductor structures for compact, tailored superconducting quantum devices.

\end{abstract}

\maketitle

\section{Introduction}
State-of-the-art transmon qubits rely on Josephson junctions (JJs) with amorphous insulating weak links. However, their large bandgaps necessitate extremely thin barriers, compromising uniformity, reproducibility, and precise control of junction critical current which controls qubit frequency and charge sensitivity. Pinholes and grain boundaries in these materials further constrain JJ sizes, and transmons based on such JJs typically require large footprint shunt capacitors for ideal transmon capacitances \cite{koch_charge-insensitive_2007}.

In contrast, semiconducting weak links offer precise control of junction properties over a wide range \cite{shim_bottom-up_2014}, and their reduced bandgaps allow for thicker barriers, which are less prone to pinhole formation compared to typical insulating barriers \cite{willsch_observation_2024}. This enables larger-area JJs with built-in capacitance, forming a smaller-footprint ``merged-element'' transmon (MET) \cite{zhao_merged-element_2020}. Recent advances in vertical JJs with semiconducting weak links have achieved fine control of junction critical current over a wide range \cite{olaya_Nb/a-Si/Nb_2023} but relied on amorphous barriers that host high densities of microwave-active two-level systems, a primary factor limiting superconducting qubits lifetimes \cite{deleon_materials_2021}.

Crystalline materials, characterized by their lack of grain boundaries, atomically pristine interfaces, and reduced two-level fluctuator densities \cite{oh_elimination_2006}, offer a promising alternative platform for high-quality quantum devices. Beyond epitaxial crystalline materials \cite{shabani_two-dimensional_2016, richardson_low-loss_2020, goswami_towards_2022}, layered van der Waals (vdW) materials, which can be peeled apart into atomically thin layers, present distinctive features to create low-loss and novel devices \cite{antony_making_2022, wang_hexagonal_2022}. For example, they may be readily encapsulated as protection against oxidation, and can be ``stacked'' together in complex \emph{heterostructures} of different vdW materials \cite{wang_one-dimensional_2013} with highly ordered internal interfaces \cite{dean_hofstadter_2013}. Their layered structure facilitates unprecedented uniformity and reproducibility compared to deposited thin films. To date, however, studies of vdW JJs have primarily focused either on vertical junctions incorporating weak links $\leq$ 6 vdW layers thick \cite{lee_ultimately_2015, island_thickness_2016, kim_strong_2017, lee_two-dimensional_2019, boix-constant_van_2021, tian_josephson_2021}, or on lateral junctions with monolayer vdW weak links \cite{du_josephson_2008, miao_premature_2009, lee_electrically_2011, ohtomo_josephson_2022, endres_current-phase_2023}.

In this work, we systematically investigate DC electronic transport in vertical vdW JJs with semiconducting weak links in the 2--12 nm thickness range, aimed at advancing a materials platform for transmon qubits. As we vary the thickness of the semiconductor, we observe a crossover from proximity-type to tunneling-type behavior around 7 atomic layers. These findings inform the optimal semiconductor weak link thicknesses for compact METs. As a proof of concept, we fabricate a prototype all-vdW MET and demonstrate its operation as a two-level system via dispersive readout.

\begin{figure}[t]
\centering
\includegraphics[width=1\columnwidth]{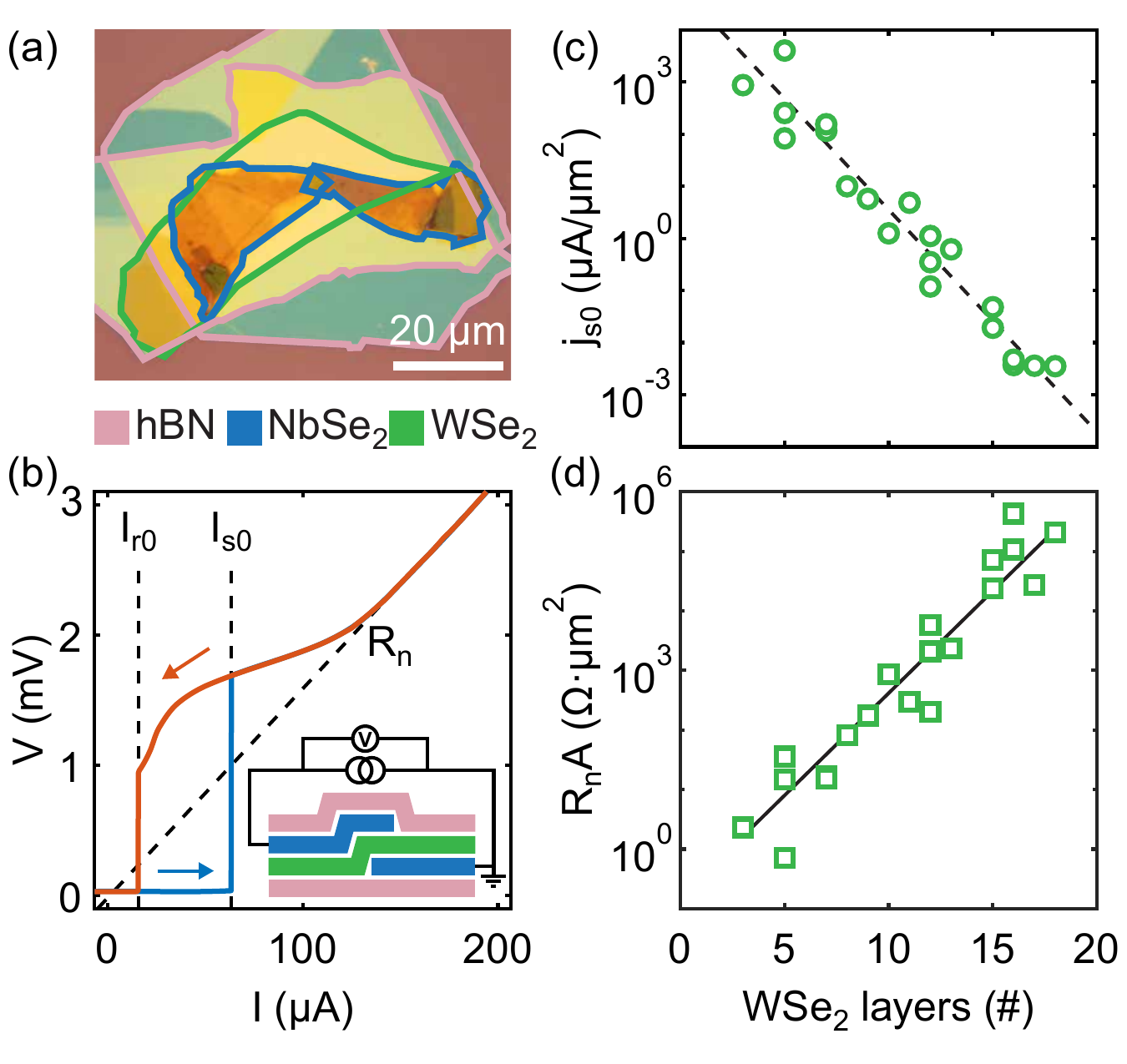}
\caption{Josephson junction characterization. (a) Optical image of an NbSe$_2$/WSe$_2$/NbSe$_2$ JJ encapsulated in hBN flakes. (b) $I$-$V$ curve of a JJ with a 9-layer-thick WSe$_2$ weak link, with dashed lines marking how $I_{s}$, $I_{r}$, and $R_n$ are obtained. Blue (red) trace indicates increasing (decreasing) bias current. (Note, the discrepancy between the retrapping current seen here and in Fig.~\ref{fig:Hysteresis}(c) is due to varying sweep rates \cite{fulton_lifetime_1974}.) Inset, schematic of a vdW JJ. (c) Switching current density at $T = 20 \text{ mK}$, $j_{s0}$, and (d) normal-state-resistance--area product, $R_nA$, as functions of WSe$_2$ weak link thickness. The black dashed line in (c) is an exponential fit to the data to guide the eye. The solid black line in (d) results from the resistivity simulations within the Thomas-Fermi model.}
\label{fig:Layer_Dependence}
\end{figure}

\section{Experimental Results}
We study vertical crystalline JJs consisting of NbSe$_2$, an anisotropic $s$-wave vdW superconductor (critical temperature $7.2 \text{ K}$, superconducting gap $1.3 \text{ meV}$ \cite{guillamon_intrinsic_2008}), sandwiching WSe$_2$, a vdW semiconductor with a $1.2 \text{ eV}$ indirect bulk bandgap \cite{kam_detailed_1982}. This is $\sim$ 6 times smaller than the bandgap of aluminum oxide and $\sim$ 5 times smaller than that of the vdW insulator hexagonal boron nitride (hBN), which acts as a tunnel barrier even in the single-layer limit \cite{britnell_electron_2012}. The lower tunnel barrier height in WSe$_2$ allows for a thicker, more uniform weak link. Consequently, the junction critical current can be controlled in small increments by changing the number of vdW layers in the WSe$_2$ weak link. We exfoliate flakes of WSe$_2$ and NbSe$_2$ from bulk crystals and build stacks with them after identifying flakes with suitable geometries (see Appendix A). We encapsulate the JJs in flakes of hBN to protect them from surface contamination and oxidation (Fig.~\ref{fig:Layer_Dependence}(a)).

In total, we study DC electronic transport in twenty JJs with WSe$_2$ thicknesses ranging from 3 to 18 layers. An example $I$-$V$ curve of a JJ with a 9-layer-thick WSe$_2$ weak link taken at temperature $T = 20 \text{ mK}$ is shown in Fig.~\ref{fig:Layer_Dependence}(b). Sweeps of increasing (blue trace) and decreasing (red trace) bias current allow us to extract key JJ parameters:~the switching current $I_{s0}$, the retrapping current $I_{r0}$, and the normal state resistance $R_n$. The subscript ``$0$'' indicates values at $T = 20 \text{ mK}$. In Fig.~\ref{fig:Layer_Dependence}(c) \& (d) we plot the switching current density $j_{s0} = I_{s0}/A$ and the product of the normal state resistance and JJ area, $R_nA$, respectively. $A$ is defined by the overlapping area of the two NbSe$_2$ flakes sandwiching the WSe$_2$ weak link. Both $j_{s0}$ and $R_nA$ obey exponential dependencies over roughly six orders of magnitude as a function of the WSe$_2$ thickness. The trends in these quantities are robust despite some scatter in the data, which we attribute to uncertainty in the JJ geometry (see Appendix A). As we will show, the trend of $j_{s0}$ versus WSe$_2$ thickness enables accurate prediction of the $0\rightarrow1$ transition energy of a transmon qubit.

\begin{figure*}[t]
\centering
\includegraphics[width=1.5\columnwidth]{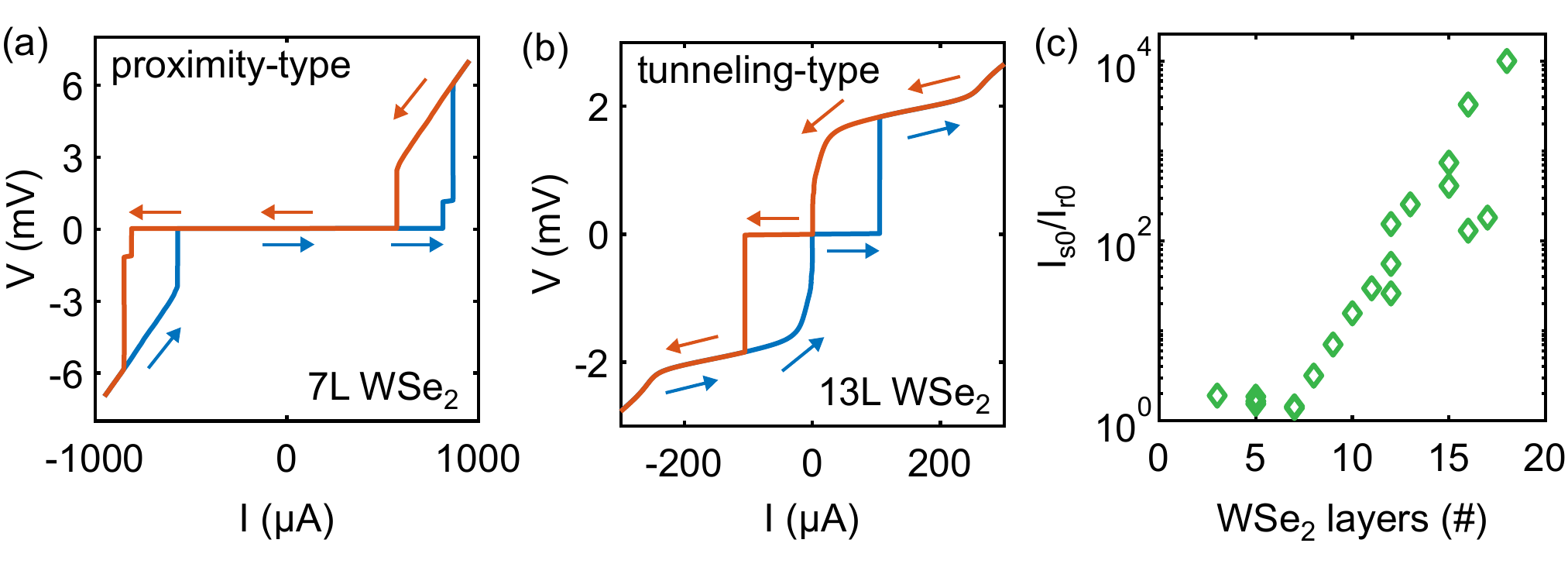}
\caption{Hysteresis in a vertical superconductor-semiconductor Josephson junction. (a), (b) DC $I$-$V$ curves for current-biased JJs with 7-layer-thick and 13-layer-thick WSe$_2$ weak links, respectively. Blue (red) traces indicate sweeps from negative to positive (positive to negative) current bias. (c) $I_{s0}/I_{r0}$ as a function of WSe$_2$ weak link thickness.}
\label{fig:Hysteresis}
\end{figure*}

To illustrate the impact of the semiconductor weak link thickness on the electronic properties of vertical JJs, in Fig.~\ref{fig:Hysteresis}(a) \& (b) we plot example DC $I$-$V$ transport characteristics of current-biased junctions with 7-layer-thick (7L) and 13-layer-thick (13L) WSe$_2$ weak links, respectively. Blue (red) traces indicate increasing (decreasing) bias current, all taken at $T = 20 \text{ mK}$. In the 13L device, the retrapping current ($I_{r0} = 0.2 \text{ {\textmu}A}$) is two orders of magnitude smaller than the switching current ($I_{s0} = 24 \text{ {\textmu}A}$) red (see Appendix A), whereas those of 7L ($574 \text{ {\textmu}A}$ and $830 \text{ {\textmu}A}$, respectively) are within a factor of two of each other \cite{Note1}. The ratio $I_{s0}/I_{r0}$ provides a measure of the JJ $I$-$V$ hysteresis, which we plot for all devices in Fig.~\ref{fig:Hysteresis}(c). In the range of 3--7 layers, $I_{s0}/I_{r0}$ is saturated around 1. Above 7 layers, the hysteresis increases exponentially with WSe$_2$ thickness. $I_{s0}/I_{r0} \approx 1$ corresponds to minimal hysteresis typical of proximity-type JJs \cite{courtois_origin_2008}. Though proximity JJs possesses a conductive weak link between the superconducting electrodes, which in principle should yield a nonhysteretic $I$-$V$, finite circuit capacitances and Joule heating in the normal state can contribute to finite hysteresis \cite{courtois_origin_2008}. On the other hand, pronounced hysteresis is characteristic of tunneling-type JJs with insulating weak links \cite{stewart_current-voltage_1968, mccumber_effect_1968}. These two different hysteretic behaviors suggest that there is a crossover from proximity- to tunneling-type JJ as the WSe$_2$ thickness increases.

\begin{figure}[b]
\centering
\includegraphics[width=0.62\columnwidth]{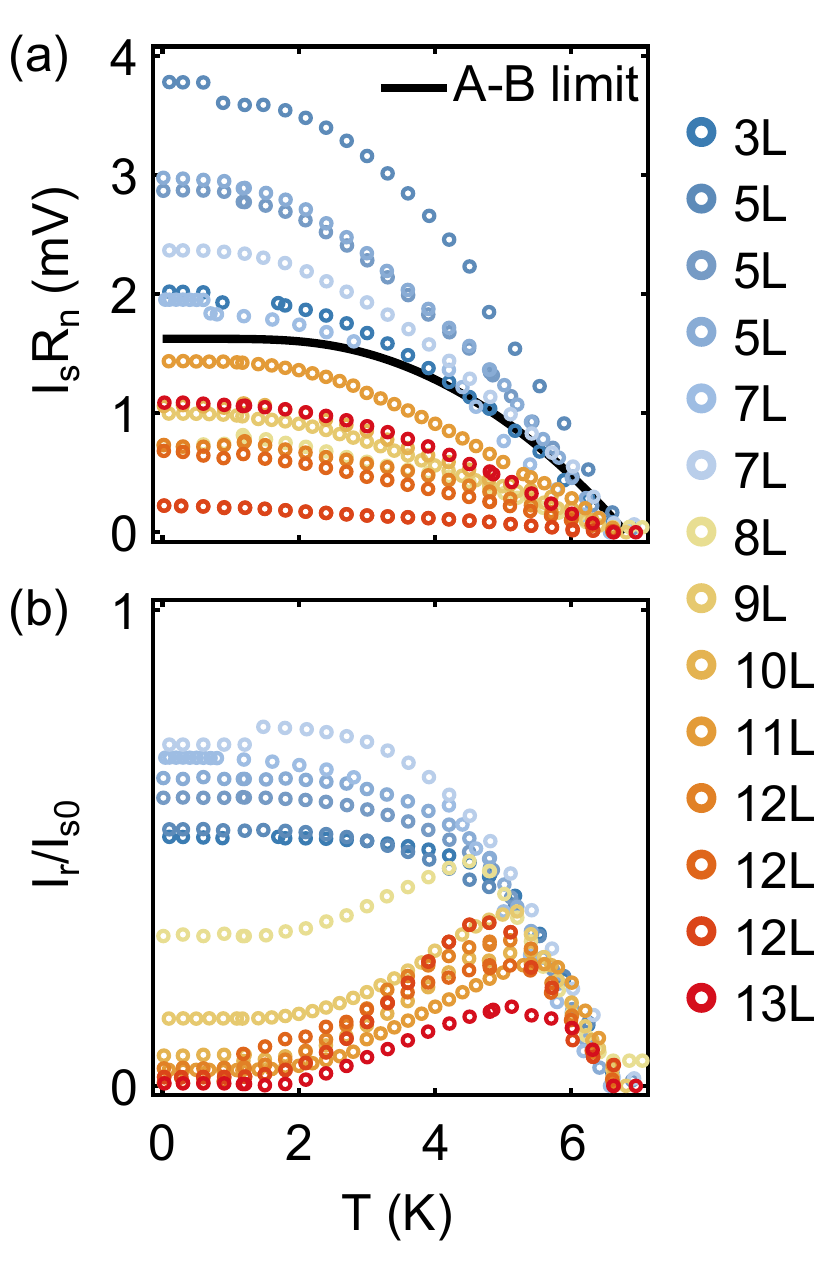}
\caption{Temperature dependence of Josephson junction properties.  (a), Temperature dependence of $I_sR_n$. (b), Temperature dependence of retrapping current $I_r$ normalized by $I_{s0}$.}
\label{fig:Temperature_Dependence}
\end{figure}

Further evidence of such a crossover is provided by the temperature dependence of the switching and retrapping currents, measured in fourteen devices with WSe$_2$ thicknesses ranging from 3 to 13 layers. In Fig.~\ref{fig:Temperature_Dependence}(a), we plot the temperature-dependent switching current $I_s$ (which we use as a proxy for the critical current $I_c$), multiplied by $R_n$. The solid black line represents the Ambegaokar-Baratoff (A-B) limit of an ideal tunnel junction:~$I_cR_n = (\pi/2)\Delta\tanh{[\Delta/2k_BT]}/e$ \cite{ambegaokar_tunneling_1963}. Here, $\Delta$ is the superconducting gap at temperature $T$, for which we use the BCS interpolation formula $\Delta = \Delta_0\tanh{[(\pi k_B T_c/\Delta_0) \sqrt{T_c/T-1}]}$ \cite{gross_anomalous_1986}. $\Delta_0 = 1.764k_BT_c \approx 1 \text{ meV}$ is the zero-temperature gap value proportional to the junction critical temperature $T_c = 6.8 \text{ K}$. Systematically, we find that the $I_sR_n$ products of all junctions with weak link thicknesses $\leq$ 7 layers (light to dark blue circles) \emph{exceed} the A-B limit, whereas those with weak links $>$ 7-layers thick (yellow to red circles) \emph{undershoot} the predicted value. The $I_cR_n$ product of a JJ exceeds the A-B limit when the transparency of the weak link is finite, \emph{i.e.},~in a proximity-type junction \cite{kulik_contribution_1975,haberkorn_theoretical_1978,golubov_current-phase_2004}. On the other hand, $I_cR_n$ in a JJ with an ideal tunnel barrier should correspond precisely to the A-B limit. However, superconducting proximity effects or a finite boundary resistance at the superconductor-weak link interface can lead to a reduction in $I_cR_n$ to a value below the A-B limit \cite{golubov_josephson_1989, golubov_josephson_1993, golubov_proximity_1995}. The grouping of $I_sR_n$ above and below the A-B limit around 7 layers of WSe$_2$ again suggests either proximity- or tunneling-type behavior occurs, depending on the WSe$_2$ thickness.

We corroborate the crossover between two Josephson junction regimes by investigating the temperature-dependent retrapping current, plotted in Fig.~\ref{fig:Temperature_Dependence}(b) normalized by the base-temperature switching current as $I_r/I_{s0}$. In JJs with WSe$_2$ weak links thinner than 8 layers (light to dark blue circles) $I_r/I_{s0}$ monotonically decreases with increasing temperature. For $\geq$ 8-layer-thick WSe$_2$ (yellow to red circles) $I_r/I_{s0}$ is nonmonotonic, wherein the retrapping current, constant at low temperatures, first rises as the temperature increases before eventually decreasing for temperatures above $5 \text{ K}$.

We may understand the nonmonotonic temperature dependence of the retrapping current in tunneling-type JJs by recalling that the current-voltage relation of a JJ is related to the evolution of the difference in the phases of the order parameter between the two superconductors. In the resistively and capacitively shunted junction (RCSJ) model \cite{stewart_current-voltage_1968, mccumber_effect_1968}, the current-biased junction is modeled as a parallel combination of an ideal Josephson element, a capacitor, and a resistor, and the phase difference across the junction behaves as a particle subject to a tilted washboard potential. The damping of the motion of this ``phase particle'' is inversely proportional to the junction resistance \cite{stewart_current-voltage_1968, mccumber_effect_1968}. In a tunneling-type JJ with low damping (high junction resistance), the particle can escape from the potential well when the current bias is increased, leading to a transition from the zero-voltage state to the resistive state. However, due to its inertia, the particle may not retrap back into the zero-voltage state unless the current is reduced significantly, creating hysteresis. At low temperatures, the damping is constant and the tunnel junction exhibits pronounced hysteresis. According to the quasiparticle tunneling model of Chen, Fisher, \& Leggett \cite{chen_return_1988}, as the temperature increases, thermal excitation facilitates quasiparticle tunneling across the junction, increasing the effective damping. This increases the retrapping current and reduces the hysteresis. 
Close to the critical temperature, the superconducting gap diminishes substantially, reducing both the critical and retrapping currents. This process explains the qualitative behavior of $I_r/I_{s0}$ in JJs with WSe$_2$ thickness $\geq$ 8 layers. In contrast, in proximity-type JJs where the weak link can readily conduct quasiparticles between the superconducting electrodes, the change in damping due to thermal activation of quasiparticles is negligible. Thus, the retrapping current in proximity-type JJs has a monotonic trend as a function of temperature, as seen for WSe$_2$ weak links thinner than 8 layers.

\begin{figure*}[t]
\centering
\includegraphics[width=1.64\columnwidth]{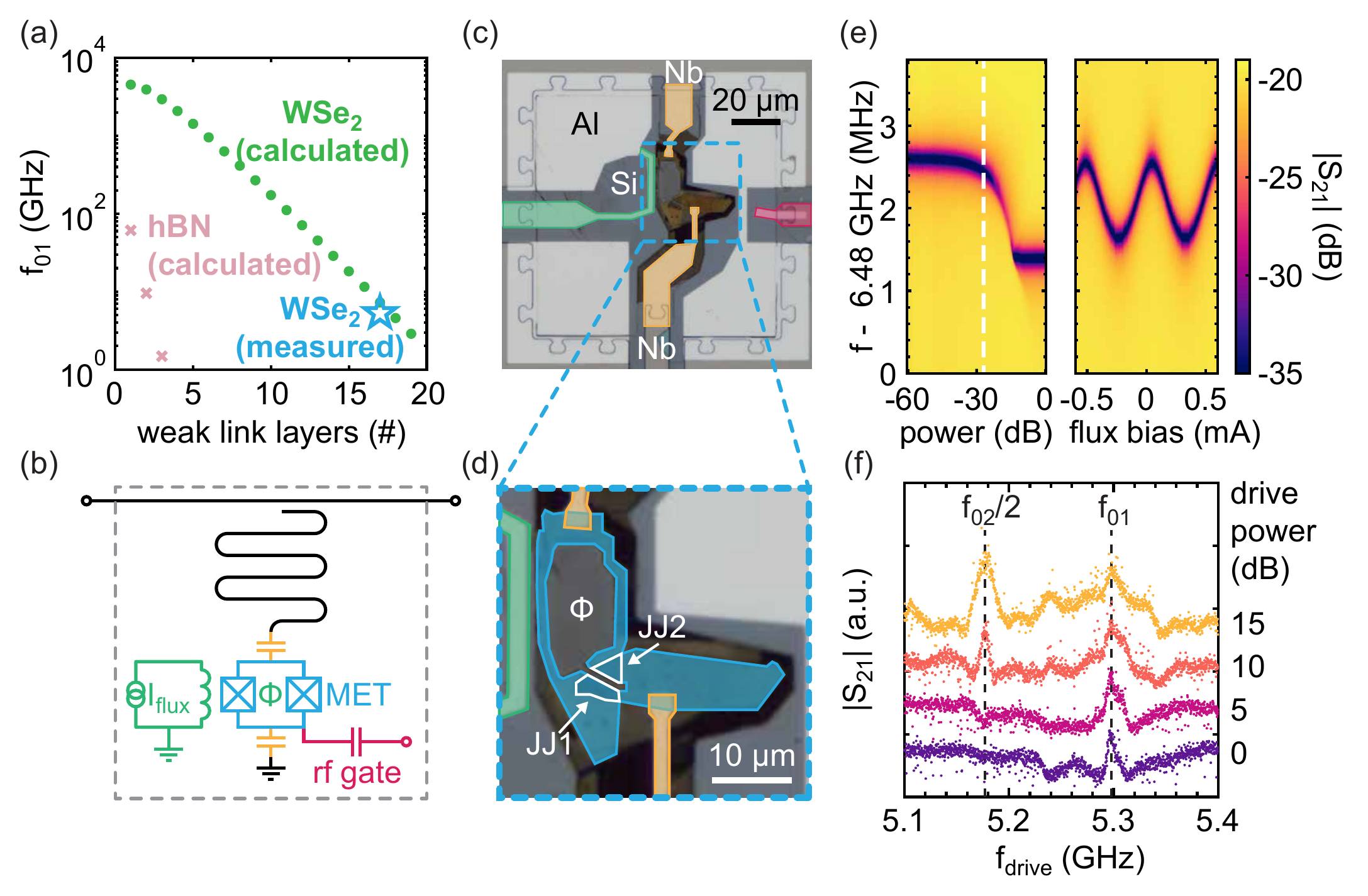}
\caption{van der Waals merged-element transmon. (a) Calculated MET 0 $\rightarrow$ 1 transition frequency $f_{01}$ for varying numbers of WSe$_2$ (green circles) and hBN (pink crosses) weak links, and the measured $f_{01}$ (blue star) for the MET shown in (c). (b) MET circuit schematic. (c) Optical image of a vdW MET. Capacitive coupling wires, flux bias line, and rf drive line are false-colored orange, green, \& red, respectively. (d) Zoom-in of the vdW stack in (c) showing the SQUID flux loop shaded blue with two JJ areas outlined in white. (e) Single-tone spectroscopy as a function of readout power (left) and flux bias current at low microwave power (right). The vertical white dashed line in the left panel indicates the power at which readout was performed in the right panel. (f) Two-tone spectroscopy. Drive pulse power increases from purple to violet to pink to gold traces, which are vertically offset for clarity.}
\label{fig:MET}
\end{figure*}

In NbSe$_2$/WSe$_2$/NbSe$_2$ JJs, a crossover from proximity-type to tunneling-type behavior with weak link thickness is plausible if we consider the relative band alignment between the two materials. \emph{Ab initio} calculations estimate a difference between the NbSe$_2$ work function and the ionization potential of WSe$_2$ of 1.07 eV, with the Fermi energy of NbSe$_2$ intersecting the valence band of WSe$_2$ \cite{shimada_work_1994, guo_band_2016}. Such a band alignment leads to an interfacial charge transfer between the two materials \cite{guan_optimizing_2017, sata_n-and-p-type_2017}, which can form an accumulation region of hole-doped WSe$_2$. 
When the WSe$_2$ weak link is thinner than the accumulation regions formed by each NbSe$_2$ electrode, the entire weak link will be doped and conductive, forming a proximity-type JJ. When the WSe$_2$ is thicker than the accumulation regions, the central layers will remain undoped, providing a tunnel barrier for the JJ. In the superconducting state, we infer from the data that $\approx$ 3 layers (2 nm) of WSe$_2$ are doped by each NbSe$_2$ electrode, leading to proximity-type behavior in JJs with WSe$_2$ weak links thinner than 7 layers, and tunneling-type behavior when the WSe$_2$ is thicker than 7 layers. While this charge transfer can give way to a superconducting proximity effect between the NbSe$_2$ and the doped WSe$_2$ layers, and subsequently the reduction in $I_cR_n$ demonstrated in Fig.~\ref{fig:Hysteresis}(c), the tunneling gap alone does not dictate the amount of band bending and proximitization. Instead, the band alignment between the two materials determines whether the Fermi energy of the superconductor will intersect the semiconductor bands or create a potential which induces band bending, as seems to be the case between NbSe$_2$ and WSe$_2$.

The relative band alignment between the two materials also dictates the strength of the exponential dependence of $j_{s0}$ and $R_nA$ on weak link thickness. We employ the Thomas-Fermi model to calculate the normal state resistance for this material system, the result of which is shown in Fig.~\ref{fig:Layer_Dependence}(d) (see Appendix B). The slope of the resistance in the semi-logarithmic plot of Fig.~\ref{fig:Layer_Dependence}(d) is determined by the Fermi level position inside the tunnel junction, which we determine to be $28.4\text{ meV}$ above the valence band of WSe$_2$ based on the fit to the data. 
Considering the individual material properties and relative band alignments, we find quantitative agreement between the measured and calculated tunneling resistance in NbSe$_2$/WSe$_2$/NbSe$_2$ JJs.

Having characterized the effects of semiconducting weak link thickness on the properties of a vdW JJ, we predict the properties of an NbSe$_2$/WSe$_2$/NbSe$_2$-based MET to determine what weak link thicknesses will result in a qubit $0\rightarrow1$ transition frequency $f_{01}$ in the range of 1--15 GHz with sufficient anharmonicity between subsequent qubit energy levels. Using the exponential fit of $j_{s0}$ from Fig.~\ref{fig:Layer_Dependence}(c) and the expected geometric junction parallel-plate capacitance, we calculate the expected MET frequencies of JJs with 1--20 layers of WSe$_2$ weak link (Fig.~\ref{fig:MET}(a)). Here, we use $f_{01} = \left(\sqrt{8E_JE_C}-E_C\right)/h$, where $h$ is Planck's constant, $E_J =\Phi_0I_c/2\pi$ is the Josephson energy with $\Phi_0 = h/2e$ the superconducting flux quantum, $e$ the elementary charge, and $E_C = e^2/2C$ the charging energy \cite{koch_charge-insensitive_2007}. For comparison, we plot the calculated $f_{01}$ for JJs with hBN weak links using previously measured tunneling resistance values \cite{britnell_electron_2012} and the Ambegaokar-Baratoff relation to estimate the critical current. Compared to WSe$_2$, hBN shows much coarser variation in $f_{01}$ with layer number due to its large bandgap, increasing by roughly an order of magnitude with each added vdW layer and requiring extremely thin weak links which are difficult to isolate and construct JJs with. In contrast, we find that JJs with 15--20-layer-thick WSe$_2$ weak links can produce METs with frequencies between 1--15 GHz.

To test these predictions, we fabricate a prototype all-vdW-material MET, comprising an NbSe$_2$/17-layer-thick WSe$_2$/NbSe$_2$ JJ. The MET readout circuit, depicted schematically in Fig.~\ref{fig:MET}(b), consists of a hanger-type microwave readout resonator capacitively coupled on one end to a microwave feedline with characteristic impedance $Z_0 = 50~\Omega$. The JJ is fashioned into a SQUID loop using reactive ion etching in a lithographically defined window and is capacitively coupled to the readout resonator at one end, and to ground at the other, by lithographically defined leads. We include additional flux bias and microwave drive lines for tuning and control. Fig.~\ref{fig:MET}(c) shows an optical image of the vdW MET, magnified in Fig.~\ref{fig:MET}(d).

In this device, we confirm dispersive interaction between the readout resonator and the MET by measuring the transmission coefficient $S_{21}$ of the feedline near the readout resonator frequency as a function of power and flux bias current (Fig.~\ref{fig:MET}(e)). We observe a distinct shift in resonator frequency below a critical power level, indicating the onset of dispersive coupling to a two-level system. At a readout power of $-27$ dB --- below this critical threshold --- the resonator frequency shows a periodic response to magnetic flux threaded through the SQUID loop, affirming that it is interacting with the frequency-tunable vdW SQUID. In Fig.~\ref{fig:MET}(f), we plot two-tone spectroscopy of the device taken at the flux-bias ``sweet spot'' where the change in the readout resonator frequency versus flux is minimized. Here, we input to the feedline a continuous weak probe tone near the readout resonator center frequency, as well as a microwave pulse at frequency $f_\text{drive}$, intended to create excitations in the two-level system. We record the change in $|S_{21}|$ at the probe frequency that results as we vary the $f_\text{drive}$ and the drive amplitude. At low amplitudes (purple data), we observe a sharp peak in the response at $f_\text{drive}\approx5.30\text{ GHz}$, which we identify as $f_{01}$. As we increase the drive amplitude (purple to violet to pink to gold) the peak at $f_{01}$ broadens while a second peak at $\approx5.18\text{ GHz}$ emerges. We designate this to be $f_{02}/2$, half the $0\rightarrow2$ transition frequency, emerging as the result of a two-photon excitation process. From this, we obtain the transmon anharmonicity $\alpha = f_{21} - f_{01} = -242 \text{ MHz}$. We calculate $E_J/E_C = 66$ for this qubit, placing it well within the transmon regime \cite{koch_charge-insensitive_2007}. The measured $f_{01}$ is within 10\% of the expected value, validating that the DC transport characterization of vdW JJs with semiconducting weak links can be used to accurately design transmon qubits.

\section{Conclusion}
In summary, we have studied DC electronic transport in crystalline vertical Josephson junctions made of superconducting NbSe$_2$ and semiconducting WSe$_2$ weak links with thicknesses varying from 3--18 vdW layers. Trends in the junction switching current, retrapping current, and hysteresis as functions of WSe$_2$ thickness and temperature reveal a crossover from proximity-type to tunneling-type junction behavior around 7 layers of WSe$_2$. We attribute this crossover to the relative band alignment between the superconductor and semiconductor. However, the Thomas-Fermi model does not predict such a crossover, raising fundamental questions about the nature of the proximitization between van der Waals superconductors and semiconductors. The observation of this crossover underscores the importance of the choice of materials used in superconductor-semiconductor quantum devices.

Following from this, we comment on how the switching and retrapping current can provide information about the internal loss of the JJ. In tunnel JJs, the magnitude of hysteresis is understood to be related to the junction quality factor, $Q$, as $ Q^* = (4/\pi)I_c/I_r \approx Q \equiv \omega_pRC $, a measure of the loss of the JJ at the junction plasma frequency $\omega_p$ (equivalent to the MET transition frequency here) \cite{stewart_current-voltage_1968}. For NbSe$_2$/WSe$_2$/NbSe$_2$ JJs, we find $Q^*$ exponentially increases with WSe$_2$ thickness up to $\approx 10^4$ for an 18-layer-thick weak link. While this quantity should only be taken as a lower bound for the junction quality factor, as Joule heating during the DC measurement may lead to a reduced hysteresis in tunnel JJs, it can serve as a proxy for comparing the quality of different materials. In Appendix C, we show that by substituting WSe$_2$ for MoS$_2$, a vdW semiconducting weak link with a similar tunnel barrier height but different band alignment, we observe less damping and higher $Q^*$ for the same thicknesses of weak link.

Using the exponential dependence of the JJ critical current density on weak link thickness, we calculated the frequencies of merged-element transmon qubits based on these crystalline JJs and validated the accuracy of these predictions by fabricating a prototype all-crystalline-material-based MET. Importantly, these qubits offer a remarkably small footprint with qubit transition frequencies that can range from a few GHz to millimeter-wave frequencies \cite{anferov_millimeter‑wave_2025}. Our results support that crystalline superconductor-semiconductor JJs are a promising platform for compact superconducting qubits with distinctive design flexibility and tunability.

\section*{ACKNOWLEDGMENTS}
This work was primarily supported by the Army Research Office under Contract W911NF-22-C-0021 (NextNEQST SuperVan-2). Synthesis and characterization of WSe$_2$ crystals (M.H.) and the use of facilities and instrumentation for sample assembly (J.B.) were supported by National Science Foundation through the Columbia University, Columbia Nano Initiative, and the Materials Research Science and Engineering Center (DMR-2011738). Theoretical modeling (V.P.) was supported by the National Science Foundation (Grant No.~2235276). Synthesis of boron nitride (K.W.~and T.T.) was supported by the Elemental Strategy Initiative conducted by the MEXT, Japan (Grant No.~JPMXP0112101001) and JSPS KAKENHI (Grant Nos.~JP19H05790 and JP20H00354). J.H.~acknowledges support from the Gordon and Betty Moore Foundation’s EPiQS Initiative, Grant GBMF10277. J.P.~acknowledges support from the education and training program of the Quantum Information Research Support Center, funded through the National Research Foundation of Korea (NRF) by the Ministry of Science and ICT (MSIT) of the Korean government (No.~2021M3H3A1036573).

\section*{APPENDIX A: Methods}
WSe$_2$ crystals are synthesized using the two-step flux synthesis method \cite{liu_two-step_2023}. Synthesis of MoS$_2$ was performed by reacting Mo (99.997\%) and S (99.999\%) (1:2+150 mg excess S) in a eutectic flux of CsCl and NaCl, as described in Ref.~\cite{cevallos_liquid_salt_2019} under vacuum ($\sim$10$^{-5}\text{ Torr}$) in a quartz ampoule. The ampoule was heated to 1000 $^\circ$C over 24 h. At peak temperature, a temperature gradient was applied with the hot end of the ampoule ($1000~^\circ\text{C}$) at the Mo and S precursor. This resulted in the crystals primarily being deposited at the cold end ($900~^\circ\text{C}$). The ampoules were then naturally cooled to room temperature by shutting the furnace off. Crystals were extracted by washing in DI water to remove the salt flux. The as-extracted crystals were then reloaded into a new quartz ampoule with excess S in a Mo:S ratio of 1:100 and heated to $900~^\circ\text{C}$ for two days and then naturally cooled. Crystals were once more extracted and excess S was removed by following the chalcogen filtering method as described in Ref.~\cite{liu_two-step_2023}.

We mechanically exfoliate hBN, WSe$_2$, and MoS$_2$ flakes from bulk crystals in air and use atomic force microscopy (AFM) to scan flakes for cleanliness and determine their thickness. While AFM is a sensitive and commonly used probe of vdW materials, the presence of physisorbed organic molecules or water on or under exfoliated flakes, air gaps, or instrumental offset can impede the accurate determination of vdW flake thickness using atomic force microscopy. This leads to a typical uncertainty of around $\pm$ 1 atomic layer \cite{kenaz_thickness_2023} for our WSe$_2$ and MoS$_2$ flakes. Other techniques to determine thickness like second harmonic generation may be used in future works to more accurately determine the layer number \cite{li_probing_2013}.

NbSe$_2$ flake exfoliation and vdW device stacking are performed in a glovebox under a N$_2$-rich atmosphere (< 0.5 ppm O$_2$ \& H$_2$O) to minimize oxidation during fabrication. During stacking, bubbles which commonly form in vdW heterostructures can lead to further uncertainty in the Josephson junction geometry, as the effective junction area defined by the overlap of the two NbSe$_2$ electrodes may be reduced by the presence of bubbles. We use an MMA/PMMA bilayer resist for e-beam lithography. CF$_4$ reactive ion etching is performed to modify the device area, and CHF$_3$ etching is used to etch regions of the encapsulating hBN and expose NbSe$_2$ for electrical contact. Before metal deposition, oxidized surface NbSe$_2$ layers are removed \emph{in situ} using argon ion-milling. Then, we deposit 3-nm-thick titanium sticking layer and 40--60 nm aluminum leads by e-beam evaporation in the same chamber \cite{antony_making_2022}.

We can control the device area within the precision of electron beam lithography in order to achieve desired critical currents or fabricate SQUID loops from a single junction. For example, for the MET device shown in Fig.~\ref{fig:MET}, we stacked a 23 {\textmu}m$^2$ JJ. We patterned a 0.5 {\textmu}m slit in the middle of the junction to split it into two equal-area junctions. A reactive ion etch comprising a combination of CF$_4$ and Ar gases was used to etch through the entire stack. Since the etch is anisotropic (predominantly vertical), the timing of the etch is not very important as long as it etches all the way through at least the top NbSe$_2$ flake. This process is reproducible and was also used to reduce the junction area of thinner WSe$_2$ weak link devices (3--7 layers), whose enormous critical current densities require very small junction areas ($\lesssim$ 1 {\textmu}m$^2$) ensure the critical current of the junction is smaller than that of the NbSe$_2$ flakes themselves. This is possible thanks to the sub-100-nm resolution of the electron beam lithography system.

We characterize our Josephson junctions (JJs) using 4-terminal DC transport measurements in a dilution refrigerator (Bluefors BF-LD400) with a base temperature below 20 mK. The bias current is swept through the junctions while the voltage difference across them is measured. This bias current is applied through a load resistor, whose resistance is much greater than the junction resistance, in series with a voltage source. 
The voltage across the junction is amplified and then read by a digital multimeter. To prevent high-frequency noise from exciting the junctions, a two-stage low-pass filter with a cutoff frequency of approximately 20 kHz is installed on all DC lines at the 4 K stage. 

Temperature control of the junctions mounted on the cold finger is achieved using a PID feedback loop implemented through  an AC resistance bridge (Lakeshore Model 372) and a 50 $\Omega$ heater mounted on the mixing chamber plate. To raise the temperature above 1.2 K, the cooling power is intentionally reduced by collecting most of $^3$He-$^4$He mixture and lowering pumping speed by turning off the turbo pump. All measurements are performed after the temperature has been stabilized for at least 20 minutes.

MET devices are mounted on the cold finger of the same dilution refrigerator with a base temperature below 20 mK. To reduce decoherence from external magnetic fields, the cold finger is enclosed in Cryoperm magnetic shielding. On the input side, the total line attenuation ranges from 70 dB to 84 dB, depending on the resonance frequency, with a 40 dB attenuator placed on the mixing chamber to further protect the qubit from thermal radiation. On the output side, a circulator is installed to block noise from external sources and prevent the input signal from reflecting back into the qubit. The output signal is amplified by low-temperature and room-temperature amplifiers. The qubit transition of the MET device, as shown in Fig.~\ref{fig:MET}(f), is characterized using a two-tone pulse measurement. A fixed DC current of 55 µA is applied to generate magnetic flux, positioning the transmon at its least susceptible point to flux noise. Both the cavity readout and qubit control pulses are fed through the input port of the transmission line.

\section*{APPENDIX B: Additional $I$-$V$ curves, description of critical and retrapping current extraction, and discussion of $I$-$V$ characteristics}
\begin{figure}[t]
\centering
\includegraphics[width=1\columnwidth]{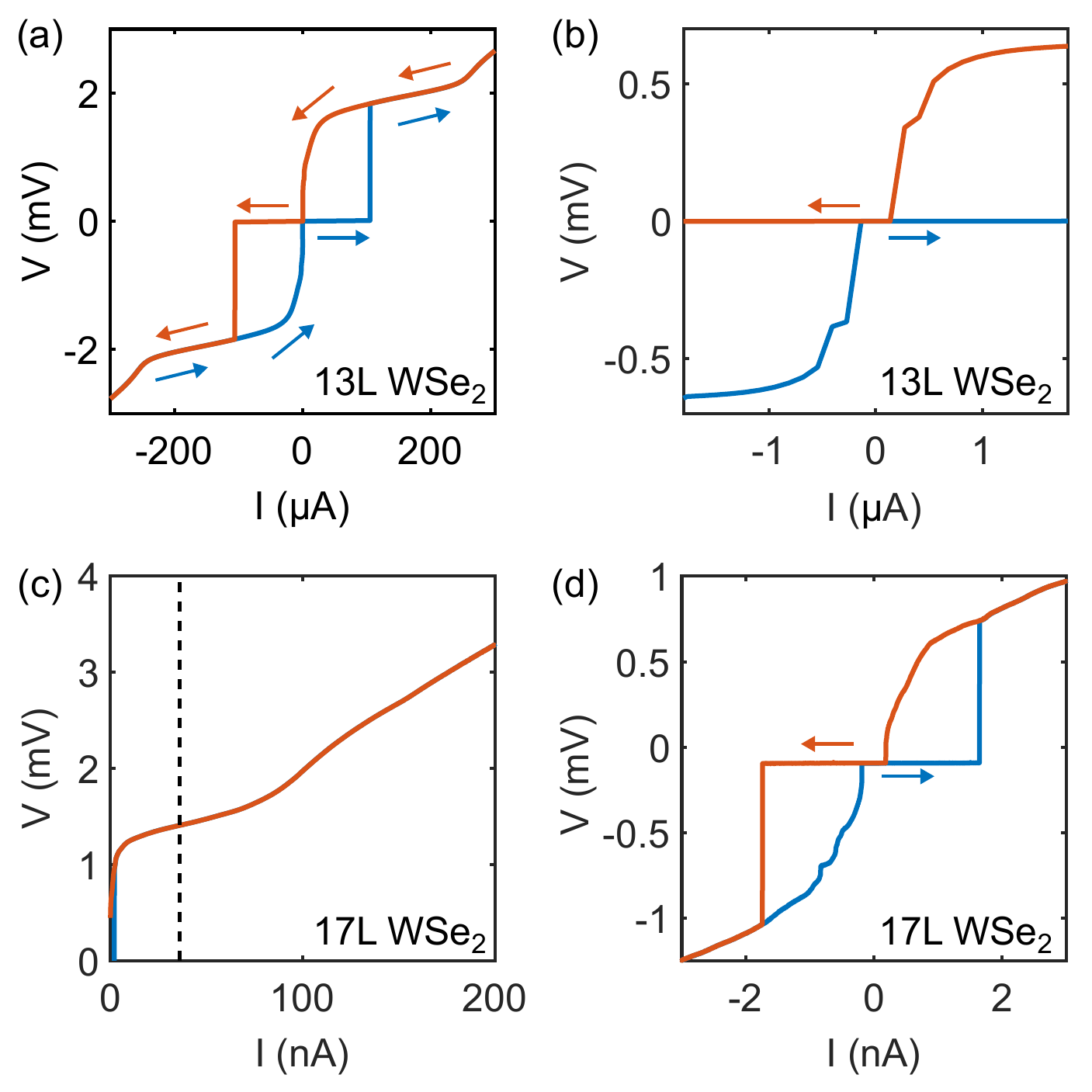}
\caption{Additional $I$-$V$ curves. (a) $I$-$V$ curve of a JJ with a 13-layer-thick WSe$_2$ weak link (same as shown in Fig.~\ref{fig:Hysteresis}. Blue (red) trace indicates increasing (decreasing) bias current sweep. (b) Zoom-in of (a) showing small retrapping current. (c) $I$-$V$ curve of a JJ with a 17-layer-thick WSe$_2$ weak link. The vertical black dashed line indicates the value of $I$ at which d$I$/d$V$ is maximized (i.e.~the superconducting gap edge), which we use as a proxy for the critical current. (d) Zoom-in of (c).}
\label{fig:SM-IV}
\end{figure}

We provide additional JJ $I$-$V$ curves to illustrate how the critical and retrapping currents are extracted in junctions with low damping and large hysteresis. In Fig.~\ref{fig:SM-IV}(a) we plot the same $I$-$V$ curve for a Josephson junction (JJ) with a 13-layer-thick WSe$_2$ weak link shown in Fig.~\ref{fig:Hysteresis}. In Fig.~\ref{fig:SM-IV}(b) we zoom in on the low-current-bias regime to show the small measured retrapping current.

In Fig.~\ref{fig:SM-IV}(c) we show an $I$-$V$ curve for a JJ with 17-layer-thick WSe$_2$. In this particular device, both the switching current and retrapping current are very small, and the former fluctuates greatly between successive measurements, making estimation of the critical current using the switching current difficult. We attribute these fluctuations to noise in our measurement system which excites the junction to the normal state at current biases well below the critical current, which junctions with smaller critical currents are naturally more susceptible to. Specifically, the noise in the system would need to rise above the potential barrier of the washboard potential, $\Delta U = 2E_J$, where $E_J = \Phi_0I_c/2\pi$ is the Josephson energy. Using the measured $I_{s0}$ of $\sim 2$ nA in place of $I_c$, this barrier would correspond to a value of $\sim 0.1$ K at zero bias.  However, by taking the point of the $I$-$V$ curve with minimal slope (i.e., the d$I$/d$V$ maximum), we can estimate the superconducting gap edge, as in a superconducting tunnel junction. Since junctions that do not suffer from premature switching due to fluctuations typically switch at the gap edge, we use the current bias at which d$I$/d$V$ is maximized as a proxy for the expected switching current $I_{s0}$ and, ultimately, the critical current in junctions that exhibit premature switching. Explicitly, these are our 14--18-layer-thick WSe$_2$ JJs, whose low critical current densities and relatively small junction areas yield low critical currents. In the case of the 17L junction, the estimated $I_{s0}$ would then become 36 nA, corresponding to a zero-bias $\Delta U$ of 1.7 K. The switching rate, $\Gamma$, is given by $\Gamma \propto \exp(\Delta U/k_B T_\text{esc})$, where $T_\text{esc}$ is the noise temperature of the system. We can model our device as being in the thermally activated regime with an extracted $C_J = 53\text{ fF}$ and measured $R_n = 2700~\Omega$. We find that, at a bias of 2 nA (corresponding to the measured $I_{s0}$), a $T_\text{esc} \approx 64 \text{ mK}$ is sufficient to switch the device at a $\Gamma \approx 0.1\text{ Hz}$, which is comparable to the integration time of our DC measurement. These details highlight the difficulty of measuring DC transport in JJs with small critical currents.

To measure the wide-range $I$-$V$ shown in Fig.~\ref{fig:SM-IV}(c) we apply a voltage across a 10 M$\Omega$ load resistor to supply the bias current. However, for the narrower range in Fig.~\ref{fig:SM-IV}(d) to observe the sub-nA retrapping current, we use a 1:1000 voltage divider to bias a 1 M$\Omega$ load resistor, giving us finer resolution in bias current. The use of a battery-powered instrumentation amplifier in the biasing circuit prevents ground loops among the instruments and reduces noise to help supply such small biases to our devices.

We also comment on the curvature observed in the subgap region of these $I$-$V$ curves, distinct from the commonly observed sharp transitions of Al/AlO$_\text{x}$/Al JJs. While a gradual retrapping such as those shown in Fig.~\ref{fig:SM-IV} can suggest finite subgap conduction, we point out that there are other factors which can influence the curvature of the subgap regime. For one, measurement of the subgap resistance of a Josephson junction can elucidate the density of states (DOS) of the superconductor near the gap edge and even in-gap states. However, we do not observe signatures of Andreev reflection or Andreev bound states, even in JJs with the thinnest WSe$_2$ weak links, implying there are minimal to no in-gap states. Additionally, proximitization at the superconductor-weak link interface can also influence the DOS near the gap edge and lead to some curvature of the $I$-$V$ in the subgap region. However, it must be remembered that Joule heating is induced in the normal state, which persists into the subgap region, not only influencing the retrapping behavior but also increasing the effective junction temperature and broadening the DOS near the gap edge. Since the superconducting gap of NbSe$_2$ ($\sim$ 1 meV) is approximately five times greater than that of Al ($\sim$ 200 {\textmu}eV), the Joule heating near the gap edge can be significantly greater in JJs with NbSe$_2$ electrodes, leading to curvature of the $I$-$V$ in this region due to thermal broadening of the DOS. For this reason, it may be unreliable to extract subgap resistances for these junctions since they can be reduced due to Joule heating.

\begin{figure}[t]
\centering
\includegraphics[width=1\columnwidth]{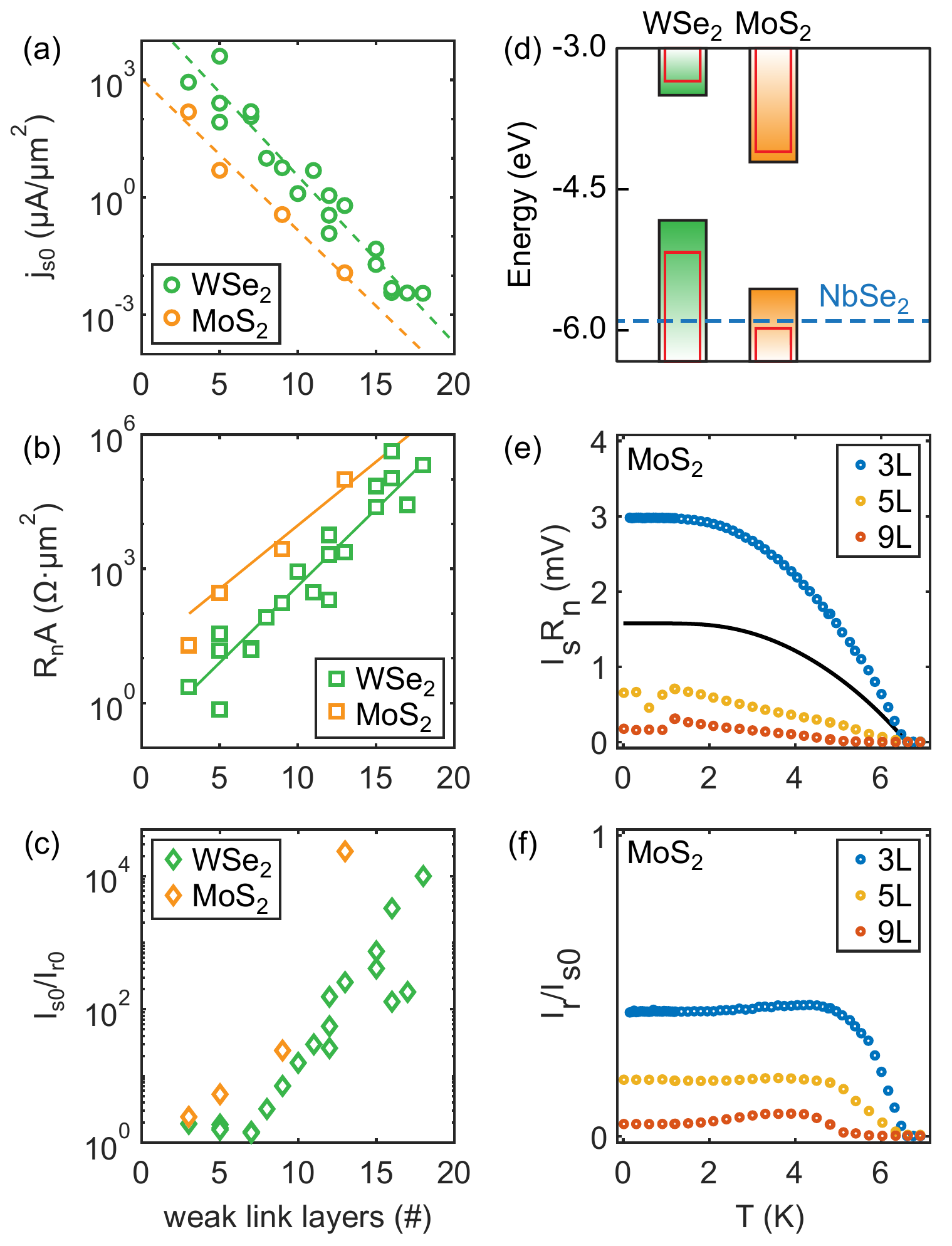}
\caption{Comparison of different semiconducting crystalline weak links.  (a), (b), (c) $I_{s0}/I_{r0}$, $j_{s0}$, and $R_nA$, respectively, versus weak link thickness for Josephson junctions with NbSe$_2$ electrodes and WSe$_2$ (green symbols) or MoS$_2$ (orange symbols) weak links. Dashed lines in (a) are fits to the data to guide the eye. Solid lines in (b) are the result of resistivity simulations within the Thomas-Fermi model. (d) Schematic band alignment of bulk WSe$_2$ (green-shaded bars) and MoS$_2$ (orange-shaded bars) relative to the NbSe$_2$ work function (blue dashed line), along with their monolayer values (red outlines) \cite{shimada_work_1994, guo_band_2016}. (e), (f) Temperature dependencies of $I_sR_n$ and $I_r$/$I_{s0}$, respectively, for JJs with MoS$_2$ weak links. Solid black line in (e) is the A-B limit.}
\label{fig:MoS2}
\end{figure}

\section*{APPENDIX C: Comparison of Different Crystalline Semiconductor Weak Links}

To emphasize the impact of the band alignment of the constituent materials on the tunneling properties and damping in a crystalline Josephson junction, we measure DC transport in vdW JJs with NbSe$_2$ electrodes and semiconducting MoS$_2$ weak links. As depicted schematically in Fig.~\ref{fig:MoS2}(d), the valence band maximum of MoS$_2$ is lower than that of WSe$_2$ \cite{guo_band_2016}, and so the band bending effect at the interface with NbSe$_2$ should be weaker when used as a JJ weak link than when using WSe$_2$. Meanwhile, the bandgaps and effective masses of the two materials are similar and so, according to the Thomas-Fermi model, their tunneling barrier heights should be roughly the same. Our DC transport results affirm this prediction, with $j_{s0}$ and $R_nA$ having nearly the same exponential dependence on weak link thickness for both WSe$_2$ and MoS$_2$ (Fig.~\ref{fig:MoS2}(a) \& (b)). However, fewer layers of MoS$_2$ are required to achieve the same tunneling properties than with WSe$_2$, suggesting that tunneling-type behavior is onset at a smaller thickness. We compare $I_{s0}/I_{r0}$ for WSe$_2$ and MoS$_2$ weak links in Fig.~\ref{fig:MoS2}(c) and find that the junctions with MoS$_2$ have lower loss for thinner weak links as indicated by a higher $Q^*$. In Fig.~\ref{fig:MoS2}(e) we plot $I_sR_n$ for JJs with 3-, 5-, and 9-layer-thick MoS$_2$ weak links, and we find that the crossover from above to below the A-B limit (black line) occurs between 3 and 5 layers, once again suggesting a crossover from proximity- to tunneling-type behavior. Meanwhile, the temperature dependence of $I_r/I_{s0}$, plotted in Fig.~\ref{fig:MoS2}(f) is less indicative of a crossover but nonmonotonicity appears to enhance with increasing layer number. Regardless, the data suggest that by changing the band alignments of the constituent JJ materials, we may change the tunneling strength sufficiently to yield different amounts of Josephson coupling and damping for the same weak link thicknesses and tunnel barrier heights.

\section*{APPENDIX D: Thomas-Fermi Model for Self-Consistent Potential in the Tunneling Region}

We employ the Thomas-Fermi model to simulate charge carrier density in layers of WSe$_2$ or MoS$_2$ between NbSe$_2$ electrodes by solving the Poisson equation for the  self-consistent potential $V(x)$,
\begin{equation}\label{Eq_Poisson}
  \epsilon\frac{d^2V(x)}{dx^2}=-\rho(x),
\end{equation}
where $\epsilon$ is the dielectric constant in the perpendicular to the semiconductor plane direction, one of two fitting parameters. The best fit of $\epsilon=5.5$ for both WSe$_2$ and MoS$_2$ is consistent with values reported in literature \cite{laturia_dielectric_2018, belete_dielectric_2018, hou_quantification_2022}. The charge carrier density $\rho(x)$ is found from the 2D carrier density on each WSe$_2$ layer $n_i$ separated by distance $d=6.5$ \AA,
\begin{equation}\label{Eq_rho}
  \rho(x)=\sum_{i=1}^N n_i \exp{\left(-(x-d\cdot i)^2/\sigma^2\right)}/\left(\sqrt{\pi}\sigma\right),
\end{equation}
where we employ Gaussian broadening with $\sigma=3 $ \AA \ \ to mimic the finite width of the electron cloud between the crystalline semiconductor layers. The 2D charge carrier density $n_i$ depends on the electrostatic potential $V_i=V(d\cdot i)$ in the middle of the monolayers, which is found self-consistently via 
\begin{equation}\label{Eq_ni}
n_i(V_i) = k_BT \cdot \frac{2m_h}{\pi \hbar^2} \cdot \ln \left( \frac{1 + e^{({-eV_i+\Delta W})/{k_BT}}}{1 + e^{({eV_i - E_g-\Delta W})/{k_BT}}} \right),
\end{equation}
where $m_h=0.53~m_e$ \cite{zhao_electronic_2019} ($0.64~m_e$ \cite{kormanyos_kp_2015}) is the hole effective mass in WSe$_2$ (MoS$_2$) in units of electron mass $m_e$, $\Delta W=1.07 \text{ eV}$ ($0.34 \text{ eV}$) \cite{shimada_work_1994, guo_band_2016} is the workfunction mismatch between NbSe$_2$ and WSe$_2$ (MoS$_2$), and $E_g$ is the bandgap of the semiconductor. We choose the Fermi energy in the metal as zero of energy and the boundary condition of $V(x=0)=V(x=L)=0$ V, where $L= d \cdot(N+1)$,  imply hole doping of the semiconductor near the contacts.

Once the self-consistent potential $V(x)$ is obtained for junctions with different numbers of layer $N$, we use WKB approximation to calculate tunneling conductance according to: 
\begin{eqnarray}\label{Eq_WKB}
G=G_M \int_{-\infty}^{+\infty} T(E) \left( -\frac{\partial f(E, E_F, T)}{\partial E} \right) dE,
\nonumber \\
T(E) = \exp \left( -2 \int_{0}^{L} \sqrt{\frac{2m_h (E - eV(x)-\Delta W)}{\hbar^2}} \, dx \right)
\end{eqnarray}
where $G_M = 1.6 \text{ S$\cdot${\textmu}m}^{-2}$ ($0.026 \text{ S$\cdot${\textmu}m}^{-2}$) is a fit parameter which has the meaning of a product of ballistic conductance of NbSe$_2$ and an additional scattering barrier at the NbSe$_2$/WSe$_2$ (NbSe$_2$/MoS$_2$) interface due to the wavefunction mismatch in two materials. The resulting resistance $1/G$ simulated at $T\rightarrow 0$ K limit is plotted in Fig.~\ref{fig:Layer_Dependence}(d) and Fig.~\ref{fig:MoS2}(b).

\bibliographystyle{apsrev4-1}

%

\end{document}